\documentstyle[aps,floats,twocolumn]{revtex}
\input BoxedEPS.tex
\SetepsfEPSFSpecial
\HideDisplacementBoxes
\begin{document}
\title{Non-destructive spatial heterodyne imaging of cold atoms}

\author{S. Kadlecek, J. Sebby, R. Newell, and T. G. Walker}
\address{Department of Physics, University of Wisconsin-Madison, Madison,
Wisconsin, 53706}
\date{\today}
\maketitle

\begin{abstract}
We demonstrate a new method for non-destructive imaging of laser-cooled
atoms.  This spatial heterodyne technique forms a phase image by interfering
a strong carrier laser beam with a weak probe beam that passes through
the cold atom cloud.  The  figure of merit equals or exceeds 
that of phase-contrast imaging, and the technique can be used over a wider range of 
spatial scales.  We show images of a dark spot MOT taken with imaging 
fluences as low as 61 pJ/cm$^2$ at a detuning of 11$\Gamma$, resulting in 0.0004 
 photons scattered per atom.
\end{abstract}

\bigskip

\narrowtext

In this paper we demonstrate a new ``spatial heterodyne'' method for non-destructive imaging
of trapped atoms.  As with other non-destructive 
techniques, spatial heterodyne imaging  minimizes the 
number of absorbed photons required for an image and is therefore particularly useful 
in applications such as Bose-Einstein Condensation\cite{BEC}, magnetic trapping, and far-off-resonance trapping  that are
particularly sensitive to heating and optical pumping from absorbed  photons.  

Off resonant, non-destructive imaging of clouds of trapped atoms \cite{Ketterle} has been 
previously demonstrated using several different methods, all of which image the phase shift produced by the atoms on a collimated
probe laser:  dark-ground imaging\cite{Andrews96}, polarization-rotation imaging \cite{Bradley97}, 
and phase-contrast imaging\cite{Andrews97}.  Non-destructive detection without 
imaging was recently demonstrated using FM spectroscopy\cite{Savalli99}. The 
most popular of these methods, the phase-contrast technique, uses a small 
($\sim 10-100$ $\mu$m) $\pi/2$ phase mask that is inserted into  the imaging 
laser focus at the Fourier plane of an imaging lens.  In the image plane the $\pi/2$
phase-shifted laser field interferes with the signal field produced by the atoms to
give an image intensity that is linear with respect to the atom-induced phase shift.

To implement spatial heterodyne imaging, we used two laser beams: a  carrier laser beam which does not pass 
through the trapped atoms, and a probe beam which is phase shifted as it passes through the atom cloud.  The beams
are coincident on a CCD camera and straightforward digital 
image processing techniques use the resulting interference pattern to reconstruct the phase shift due to the cloud.

Spatial heterodyne imaging has several practical advantages for non-destructive 
imaging.  First, there is no need for precision fabrication and alignment of a 
phase plate.  Second, it has a significant signal-to-noise advantage for low imaging
intensities.  Third, at high intensities it has a larger signal per absorbed photon, allowing   the large dynamic range
of CCD cameras to be better  used.  Fourth, the method works over a wide range of spatial scales.
Finally, rejection of spurious interference fringes due to various
optical elements such as vacuum windows is automatically accomplished.

\begin{figure}[bh]
\BoxedEPSF{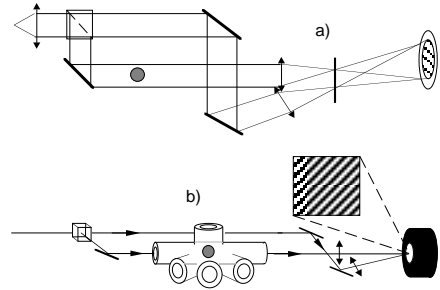 scaled 530}
\caption{Apparatus.}\label{geom}
\end{figure}

The principle of spatial heterodyne imaging is similar to heterodyne
spectroscopy \cite{Yariv76}, with interference occuring in the spatial rather than the temporal domain.  As shown in
Fig.~\ref{geom}a) a probe beam  of intensity $I_p$  travels through a cloud of trapped atoms and accumulates a 
position dependent phase shift $\phi({\bf r})$ due to the index of refraction 
of the atoms. A lens placed in this beam images the atom cloud onto a CCD detector.  A  carrier 
beam of intensity $I_c$, and derived from the same laser as the probe beam 
interferes with the probe beam at an angle $\theta$.  For convenience, we 
assume equal radii of curvature for the carrier and probe beams. The 
interference pattern on the CCD detector $I({\bf r})$ is a set of straight line 
fringes whose position is determined by an overall phase shift between the 
beams $\chi$, and which are distorted by the accumulated phase shift from the atoms:
\begin{equation}
I({\bf r})=I_c+I_p+2\sqrt{I_c I_p}\cos({\chi+2\pi\theta {\bf \hat{k}_\perp\cdot r}/\lambda-\phi({\bf r})})
\label{fringe}
\end{equation}
from which $\phi({\bf r})$ can be reconstructed. Here $\hat{k}_\perp$ is a unit vector pointing along the direction of
the component of the carrier wavevector ${\bf k}$ perpendicular to the direction of the probe beam.

The phase shift $\phi$ is most easily determined in two limits: $\theta\ll
\delta /\lambda$ (parallel mode) and $\theta\gg\delta /\lambda$ (tilted mode), where $\delta $ is the
desired resolution element on the image.  In the parallel mode the phase of
the interference pattern is uniform across the cloud image, and the resulting
interference pattern is (with $\chi=\pi/2$):
\begin{equation}
I({\bf r})=I_c+I_p+2\sqrt{I_c I_p}\sin{\phi({\bf r})}
\end{equation}
If $I_c=I_p$, this is identical to phase contrast imaging.  If not, the signal size is increased by a factor of $\sqrt{I_c/I_p}$.
  The spatial variation of the phase
shift from the cloud becomes a spatial variation of the intensity at the CCD
detector, producing a real image on the detector.  In practice the phase shift $\chi$ between the two 
beams must be stabilized using feedback.

For this paper we have implemented spatial heterodyne imaging in the tilted mode.
In this case a set of high spatial frequency fringes appear and the effect 
of the atom cloud is to give a spatially varying phase shift to these fringes.
The analysis of the fringes then proceeds in a manner highly analogous to 
lock-in detection: we demodulate the interference pattern to zero spatial 
frequency and apply a low-pass filter to the result.  FFT techniques make the demodulation and filtering 
 efficient (2 sec for a $784\times520$ pixel camera on a 400 MHz Pentium).  It is not necessary to stabilize the relative
phase between the probe and carrier beams.

To demonstrate the method we use an atom cloud with an on-resonant optical 
thickness of about 15 in a dark spot $^{87}$Rb MOT 
\cite{Ketterle93,Anderson94}. The experimental arrangement is shown in
Fig.~\ref{geom}b.  Typically $3\times 10^7$ atoms from a MOT are accumulated in the
dark state at a density of roughly $5\times 10^{11}$ cm$^{-3}$ by imaging a 1 mm obstruction in the MOT repumping laser onto
the trap.  The atoms in the dark spot are quite sensitive to  resonant light
and hence absorption imaging is difficult.  The imaging laser beam is
tuned in the range of $ 2-11\Gamma$ away from the 
$^{87}$Rb 5S$_{1/2}$(F=1)$\rightarrow$5P$_{1/2}$(F=2) resonance, switched via an acousto-optic modulator, and then split
into  two beams by a non-polarizing beamsplitter.  The probe beam is attenuated by a factor of 1-200 by
a neutral density filter before passing through the atom cloud, which is imaged 
onto a CCD array. An interference filter placed in the Fourier plane of the imaging lens rejects 780 nm fluorescence from the
bright state trapped atoms. The carrier passes around the vacuum chamber and is incident on the  CCD detector tilted at an angle
$\theta\approx 1
\deg$.  For convenience, we  roughly match the radii of curvature of the probe and carrier beams at the 
CCD, to produce nearly straight fringes.  We tilt the fringes at an angle of typically 30$^\circ$ from the
rows of the CCD chip to avoid aliasing.

Two competing factors determine the optimum angle $\theta$.  As with 
lock-in detection, it is important to modulate the signal at somewhat higher 
spatial frequency than the smallest feature to be resolved.  The 
finite camera pixel size sets an upper limit on the modulation frequency 
without loss of fringe contrast.  We find that a fringe spacing of 4-5 pixels 
is a good compromise between resolution and fringe contrast.  In the parallel mode the full resolution of the camera is
acheived.

\begin{figure}[bht]
\BoxedEPSF{ 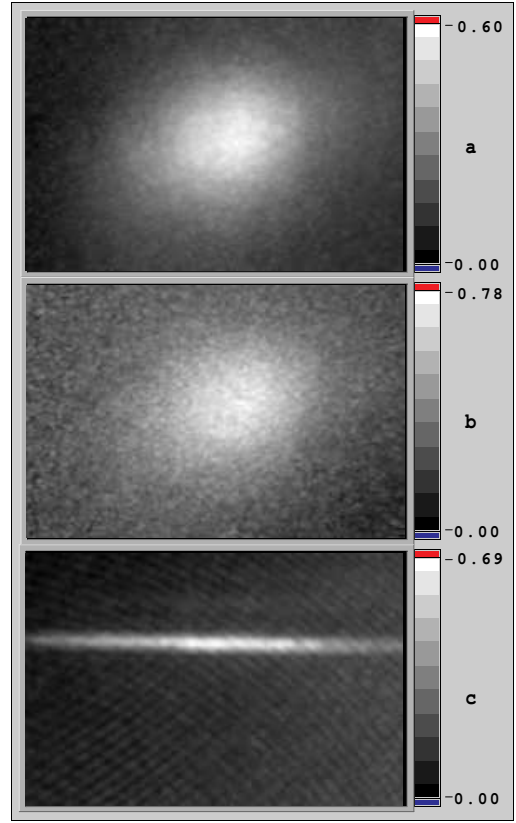 scaled 625}
\caption{a)  Image of a dark spot MOT taken at a probe
detuning $\Delta=-11\Gamma$ and carrier-to-probe intensity ratio $r=20$.  b)  Image optimized for minimum light scattering:
$\Delta=-11\Gamma$, $r=60:1$.  This image required approximately 0.0004 photons to be scattered per atom.  c)  Side-on image
of a dark sheet trap formed by  an 59 $\mu$m wire image in the repumping beam.  The image of the cloud is approximately 51
$\mu$m. The right-hand scale is the phase shift in radians. }\label{data}
\end{figure}

To begin processing we subtract off reference images of each laser beam.  This leaves only the interference term in
Eq.~\ref{fringe},  which we Fourier transform.  The transform contains the phase image 
information in two sections centered on spatial wavenumbers 
$k_0=\pm 2\pi\theta/\lambda$.  We shift one of these sections to 
zero spatial frequency and attenuate the high frequencies with a filter,
typically the Gaussian filter exp(-$(3k/k_0)^2$). Finally, we take the
inverse transform whose phase (tan$^{-1}$(Im/Re)) is $\phi({\bf r})$.  This procedure 
automatically reduces spurious interference fringes that arise from various 
optical elements since they are likely to be at the wrong spatial frequency.  
To compensate for slight curvature of the interference fringes, we subtract $\phi({\bf r})$ from another 
image, similarly processed, but taken in the absence of atoms.  
This also reduces distortion due to spatial inhomogeneities in $I_c$ and 
$I_p$.

Figure~\ref{data} shows several images $\phi({\bf r})$ taken using the above procedure. At a typical line-center optical
thickness of 3-15 we have successfully imaged the dark spot trap  for a variety of detunings and carrier-to-probe intensity
ratios
$r$. Fig.~\ref{data}a) shows a typical image with $r=20$, $\Delta=-11\Gamma$, and about $1.2\times 10^{-3}$ scattered photons
per atom.  As another example, Fig.~\ref{data}b) shows an image taken at
$\Delta=-11\Gamma$ and $r= 60$.  The total fluence used to make the image was only 61
pJ/cm$^2$, corresponding to 0.0004 photons scattered per atom.  The S/N ratio on a given resolution element is
about 10 for this image.

Depending on the details of the imaging system, filtering of the Fourier
transform may limit the spatial resolution of the final image.  In our
system, with a magnification of 5 and a CCD pixel spacing of 8.8$\mu$m,
the resolution is limited to about 20$\mu$m, compared to a theoretical
diffraction limit of about 5$\mu$m.  Fig.~\ref{data} shows an image of a 50 $\mu m$ wide trap.


Depending on the application, the figure of merit for spatial heterodyne imaging is
comparable with or superior to phase constrast imaging.
For simplicity, we consider here the parallel mode.  The intensity pattern is $I_c+I_p+2\sqrt{I_c I_p}\cos(\chi-\phi(x))$, 
$I_c$   and $I_p$ are measured in numbers of photons. For small phase shifts and $\chi\approx
\pi/2$, the signal size is approximately $2\eta\sqrt{I_c
I_p}\phi$ where
$\eta$ is the quantum efficiency of the detector, typically $\sim 0.3$ for 
CCD chips in the near infrared.  Noise sources include shot noise and other sources of technical noise, $b$, such as the camera
read noise and finite resolution of the camera's A/D converter.
 The signal-to-noise ratio is therefore
\begin{equation}
(S/N)_{SH}={2\eta\sqrt{I_c I_p}\phi\over\sqrt{\eta(I_c+I_p)+b^2}}
\end{equation}
The maximum $S/N$ occurs for $I_c\gg I_p,b^2/\eta$, giving \begin{equation}
(S/N)_{max}={2\phi\sqrt{\eta I_p}}
\end{equation} which shows that there is a
minimum number of photons that must be scattered from the atoms  to acheive a given S/N. A similar relation holds for phase
contrast imaging.

A natural figure of merit for non-destructive imaging is the number of absorbed
photons required to attain the desired signal to noise ratio.  For optically
thick clouds this number is greatly reduced because the probing can be
done at large detuning\cite{Ketterle}. Thus the shot-noise limited figure of merit for
either
technique is
\begin{equation}
{S/N\over A}={2\phi\sqrt{\eta I_p}\over \alpha I_p}\approx {2\Delta\over
\Gamma}\sqrt{\eta \over I_p}
\end{equation}
for $\Delta\gg\Gamma$.

\begin{figure}[htbp]
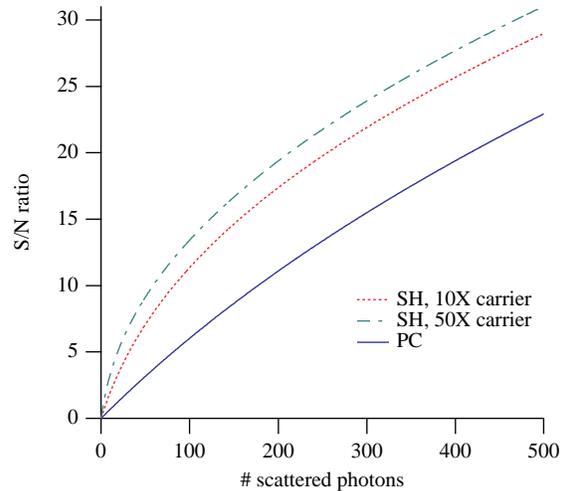

\BoxedEPSF{fom.epsf scaled 500}
\caption{S/N comparison for phase contrast imaging and spatial heterodyne
imaging.  The camera read noise is assumed to be $b=25$ e$^-$, and the
quantum
efficiency is $\eta=0.5$.}\label{fomfig}
\end{figure}

When the technical noise $b$ is significant, however,  spatial heterodyne
imaging has a better S/N ratio than phase
contrast imaging.  Figure~\ref{fomfig} compares the S/N ratio per radian of phase
shift for the two techniques.   As with any
heterodyne method, the interference between the carrier and the signal
boosts the signal level at a given probe
intensity.

Furthermore, for highest quality images it is desirable to maximize signal size and thereby minimize the discretization errors
from the A/D converter.  In this case the spatial heterodyne method offers a $\sqrt{r}$ performance enhancement as
compared to the phase contrast technique.  Fig.~\ref{data}b) shows an image taken with $r=60$, representing 3 bits of increased
signal size for fixed absorption.

We have demonstrated the spatial heterodyne method for non-destructive imaging of trapped atoms and shown that it has some
advantages over other techniques.  Our method is   a special case of of a more general class of holographic imaging techniques
that could be used with cold atoms.

Support for this research came from the NSF. We acknowledge helpful
communications with D. Jin, W. Ketterle, and S. Rolston, and
 assistance from N. Harrison.

\end{document}